\documentclass[runningheads]{llncs}
\usepackage{graphicx}
\usepackage{subfigure}
\usepackage{booktabs}
\usepackage{multirow}
\usepackage{amssymb}
\usepackage{amsmath}
\usepackage[colorlinks]{hyperref}
\usepackage{color}

\usepackage{floatrow}
\newfloatcommand{capbtabbox}{table}[][\linewidth]
\begin{document}
\title{Semi-supervised Cardiac Image Segmentation via Label Propagation and Style Transfer}
%
%
\author{Yao Zhang\inst{1, 2} \and Jiawei Yang\inst{3} \and Feng Hou\inst{1, 2} \and Yang Liu\inst{1, 2} \and Yixin Wang\inst{1, 2} \and Jiang Tian\inst{4} \and Cheng Zhong\inst{4} \and Yang Zhang\inst{4} \and Zhiqiang He\inst{5}}
\authorrunning{Zhang et al.}
%
\institute{Institute of Computing Technology, Chinese Academy of Sciences, Beijing, China \and University of Chinese Academy of Sciences, Beijing, China \and Electrical and Computer Engineering, University of California, Los Angeles, USA \and Lenovo Research, Beijing, China \and Lenovo Ltd., Beijing, China\\
\email{zhangyao215@mails.ucas.ac.cn}
}
\maketitle              
\begin{abstract}
Accurate segmentation of cardiac structures can assist doctors to diagnose diseases, and to improve treatment planning, which is highly demanded in the clinical practice. However, the shortage of annotation and the variance of the data among different vendors and medical centers restrict the performance of advanced deep learning methods. In this work, we present a fully automatic method to segment cardiac structures including the left (LV) and right ventricle (RV) blood pools, as well as for the left ventricular myocardium (MYO) in MRI volumes. Specifically, we design a semi-supervised learning method to leverage unlabelled MRI sequence timeframes by label propagation. Then we exploit style transfer to reduce the variance among different centers and vendors for more robust cardiac image segmentation. We evaluate our method in the M\&Ms challenge~\footnote{https://www.ub.edu/mnms/}, ranking $2$nd place among $14$ competitive teams. The code is available at \href{https://github.com/YaoZhang93/Semi-supervised-Cardiac-Image-Segmentation-via-Label-Propagation-and-Style-Transfer}{https://github.com/YaoZhang93/Semi-supervised-Cardiac-Image-Segmentation-via-Label-Propagation-and-Style-Transfer}.

\end{abstract}
\section{Introduction}
Cardiac disease is one of the leading threats to human health and causes massive death every year. In the clinical routine, advanced medical imaging techniques (such as MRI, CT, ultrasound) are used for the diagnosis of cardiac disease. Accurate segmentation of cardiac structures from medical images is an essential step to quantatively evaluate cardiac function and improve therapy planning, which is highly demanded in the clinical practice. 

In the recent years, deep learning models have been widely used for cardiac image segmentation and achieved promising results~\cite{chen2020deep}. However, these models could be failed on unseen datasets acquired from distinct clinical centers or medical imaging scanners~\cite{tao2019deep}. The M\&Ms challenge is motivated to contribute to the effort of building generalizable models that can be applied consistently across clinical centers~\cite{victor_m_campello_2020_3715890}. In this challenge, the cohort is composed of 350 patients with hypertrophic and dilated cardiomyopathies as well as healthy subjects. All subjects were scanned in clinical centers in three different countries (Spain, Germany and Canada) using four different magnetic resonance scanner vendors (Siemens, General Electric, Philips and Canon). The variance of data among multiple centers and vendors poses extreme challenge to the generalization of machine/deep learning models. 

\begin{figure}[!tp]
\centering
\includegraphics[width=\linewidth]{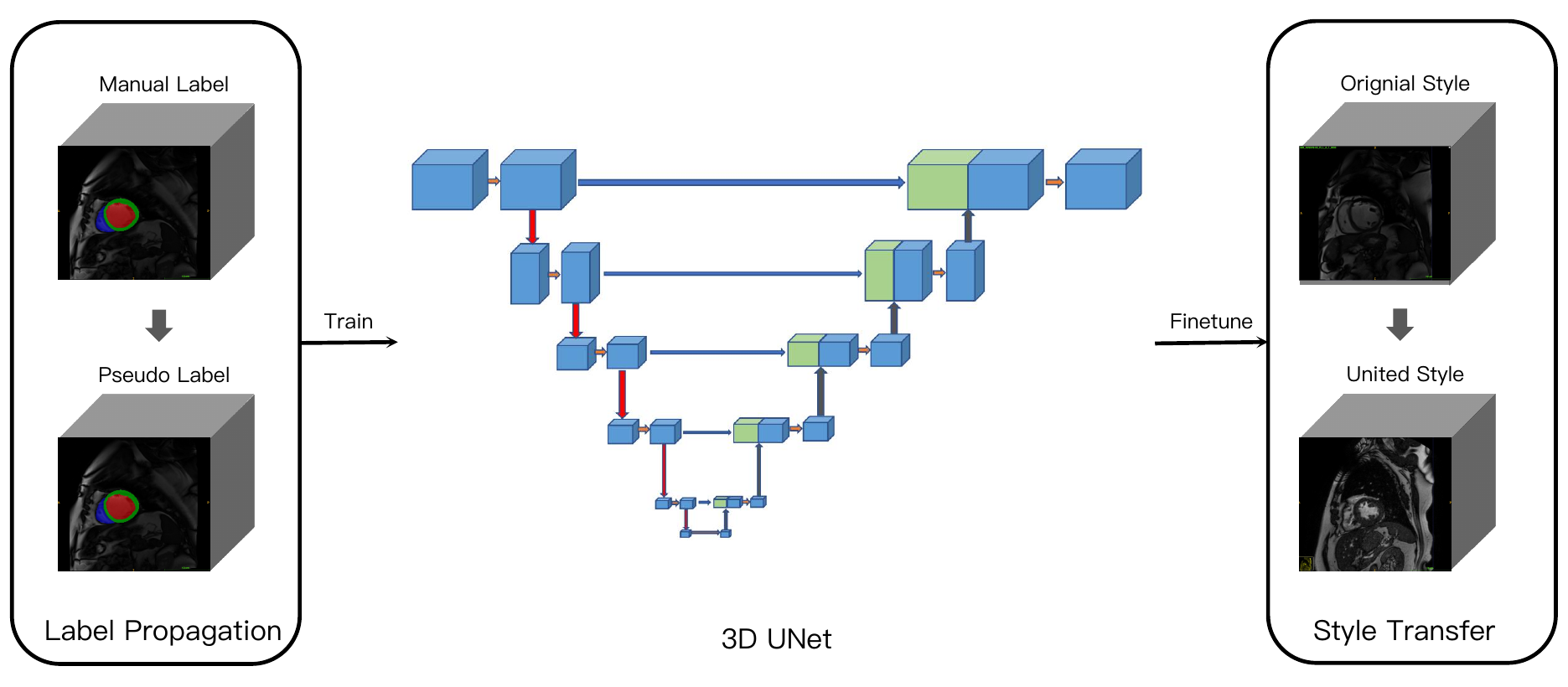}	
\caption{The pipeline of our method.}
\label{overview}
\end{figure}

The success of these deep learning models usually relies on large-scale manually annotated datasets. It requires expensive resources to manually label the images. In contrast to 3D MRI or CT images for other organs, cardiac MRI sequence timeframes has 4 dimensions (i.e., height, depth, width, and time series), which leads to much more workload for annotation. In M\&Ms challenge, only the End of Systole (ES) and the End of Diastole (ED) of the cardiac MRI sequence timeframes are annotated. Plenty of images between ES and ED stay unlabelled, which limits the performance of supervised learning.

In this paper, we propose to leverage unlabelled MRI sequence timeframes to improve cardiac segmentation by label propagation. Then we exploit style transfer to reduce the variance among different centers and vendors for more robust cardiac image segmentation.

\section{Method}
In this section, we will describe our method in detail. We employ a semi-supervised method to achieve effective cardiac image segmentation using unlabelled time-frames from the MRI sequence timeframes. The pipeline of our method is illustrated in Fig.~\ref{overview}. In contrast to~\cite{chen2019unsupervised}, we exploit label propagation to leverage unlabelled cardiac sequence. Firstly, a set of pseudo labels of the unlabelled images between ES and ED are generated. Secondly, a 3D UNet~\cite{ronneberger2015u} is trained on the data with both manual and pseudo labels. At last, as those pseudo labels is not as accurate as manual ones, the trained UNet is fine-tuned on the manually labelled data. Furthermore, in order to reduce the gap of the data from different vendors, we augment the training data through style transfer to improve the generalization of semi-supervised learning.

\subsection{Label Propagation}
Here we augment our training set from unlabeled time frames by leveraging the insights of label propagation. Label propagation is a family of semi-supervised algorithms. It advocates the exploitation of similarity between labeled and unlabeled data, granting us the ability to assign labels to previously unlabeled data.

Fortunately, this task is a well-formulated problem for label propagation for its subject-level time-series prosperity. Firstly, the annotations of End of Systole (ES) and End of Diastole (ED) are good priors as they capture the most extreme scenarios and thus can cover the transition frames between ES and ED. Meanwhile, data from different time frames within a patient share almost identical distribution, which will alleviate propagation errors caused by inter-subject variabilities. Hence, with ES and ED frames as priors, label propagation algorithms can propagate their labels to non-ES and non-ED time frames in a intra-subject manner. By doing so, the propagated labels are better constrained with specificity compared with inter-subject propagation.

\begin{figure}[!tp]
\centering
\includegraphics[width=\linewidth]{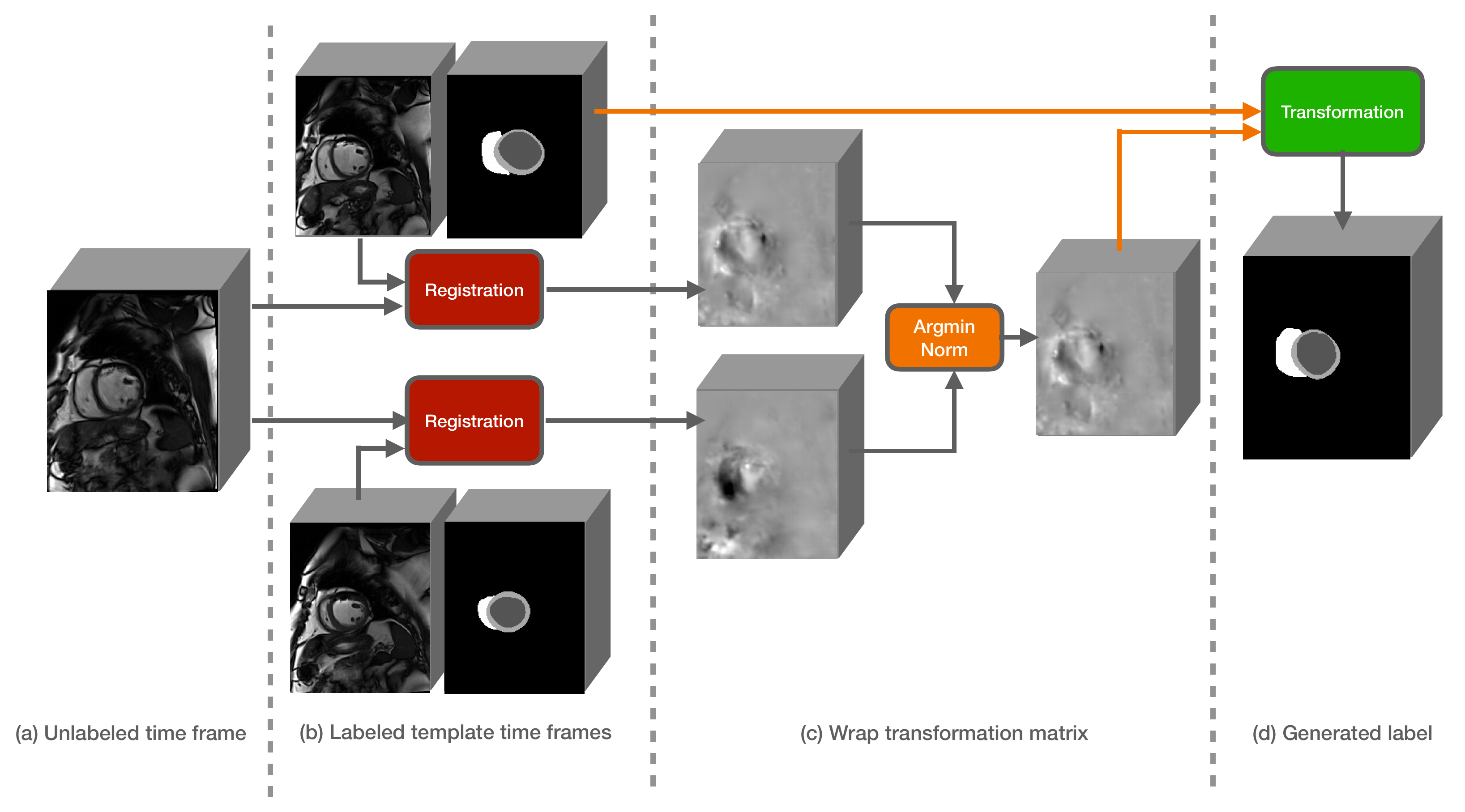}
\caption{The procedure of label propagation.}
\label{label}
\end{figure}

\begin{figure}[!tp]
\centering
\includegraphics[width=12cm, height=4cm]{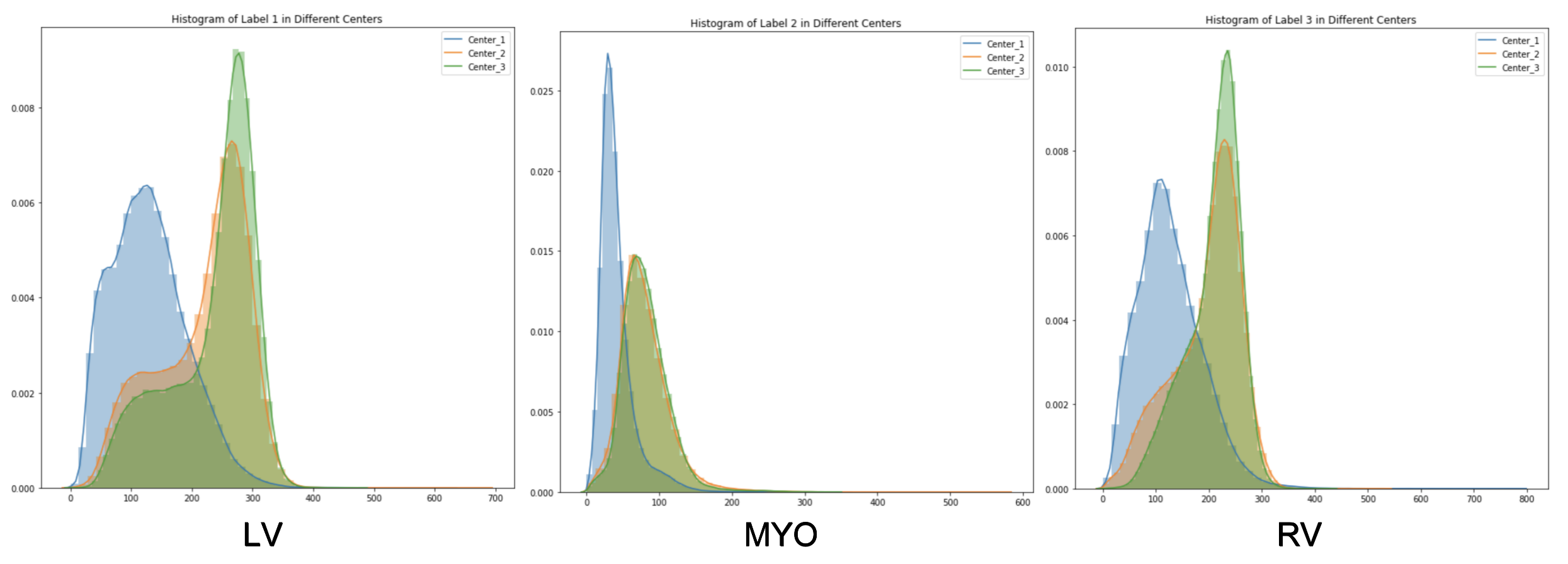}
\caption{Histogram distribution of MRI from different clinical centers. Center 1, center 2, and center 3 are marked in blue, orange, and green, respectively. Note that MRI scanners used in both center 2 and center 3 is from the same vendor, while that in center 1 is different.}
\label{hist}
\end{figure}

Specifically, we use registration algorithm to propagate labels, as shown in Fig.~\ref{label}. For each patient, the ES and ED frames are the template frames. Given a target time frame, two template frames are first registered to $T$, resulting two \textit{Warp} matrix, called $ES_{wrap}$ and $ED_{wrap}$. Naturally, we want the registered label smooth and not elasticizing from the template too much. So we then compare the norm of two $ES_{wrap}$ and $ED_{wrap}$ and choose the one with smaller norm as $T_{wrap}$. Finally, the generated label is obtained by applying the $T_{warp}$ to the corresponding template label.

\subsection{Style Transfer}
Instead of involving a complicated deep learning network for style transfer~\cite{chen2020leveraging}, we utilize a simple yet effective method, histogram matching, to achieve this. Histogram matching is a process where a time series, image, or higher dimension scalar data is modified such that its histogram matches that of reference dataset. A common application of this is to match the images from two sensors with slightly different responses, or from a sensor whose response changes. We analyze the histogram distribution between different vendors and medical centers, and find that the histogram difference between vendors is much more remarkable while that between centers is insignificant (see Fig.~\ref{hist}). Herein, we apply it to match MRI images from different vendors.

The procedure is as follows. The cumulative histogram is computed for each dataset from a vendor. For any particular value $x_i$ in the data to be adjusted has a cumulative histogram value given by $S(x_i)$. This in turn is the cumulative distribution value in the reference dataset, denoted as $T(x_j)$. The input data value $x_i$ is replaced by $x_j$.

Specifically, we sample 100 volumes from training set and then randomly select a slice from each volume as the unified reference data. Then other data is matched to the reference one. As large scale dataset benefits the deep learning method, we also augment the training data by transferring the data from one vendor to another (i.e., from vendor A to vendor B or vice versa).

\begin{figure}[!tp]
\centering
\includegraphics[width=9cm, height=7cm]{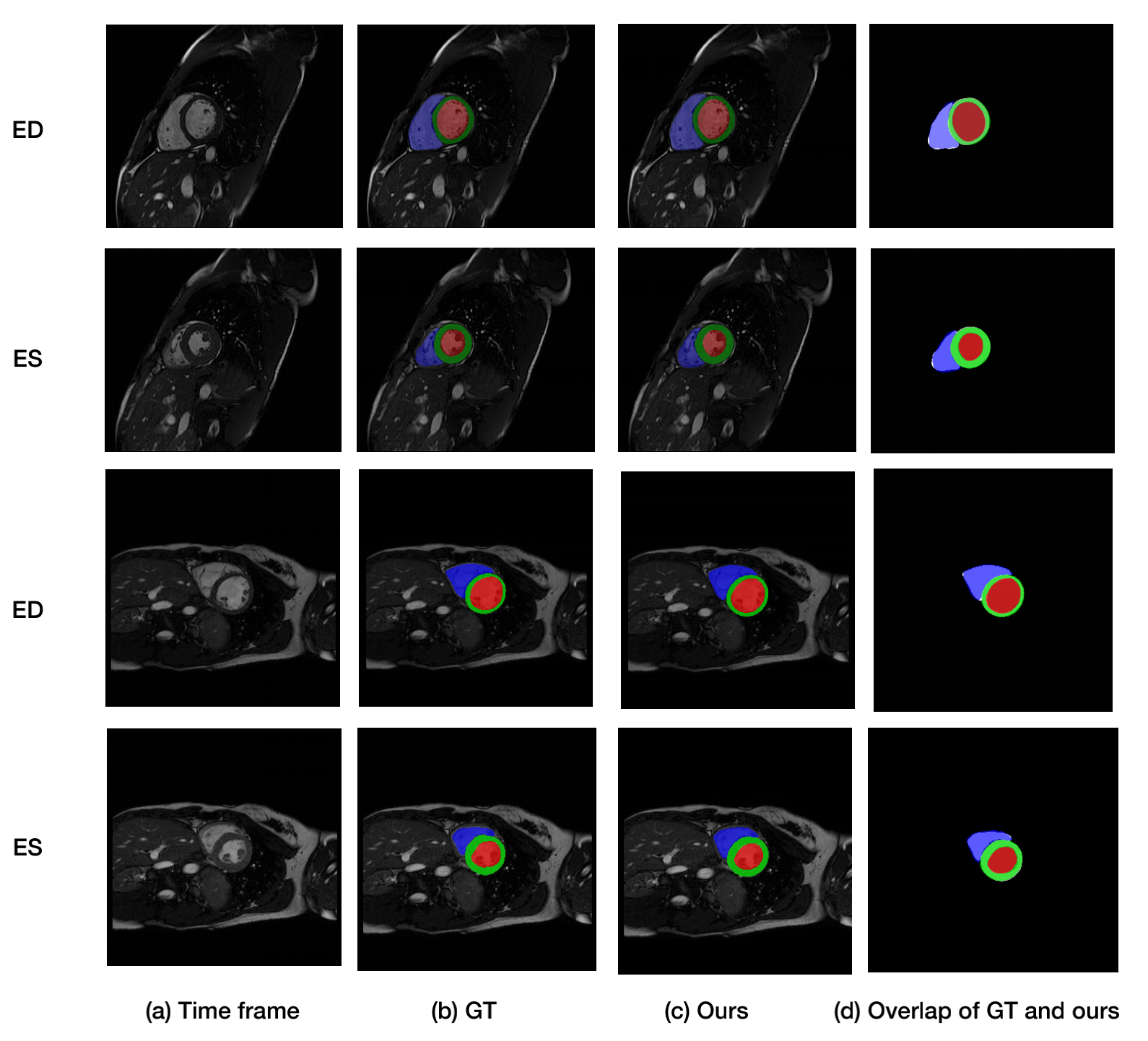}
\caption{Visual examples of accurate segmentation. From left to right, each column shows the timeframe, ground truth, prediction of the proposed approach, and the difference between ground truth and prediction. LV, MYO, and RV are marked in red, green, and blue, respectively.}
\label{good}
\end{figure}

\begin{figure}[!tp]
\centering
\includegraphics[width=9cm, height=6cm]{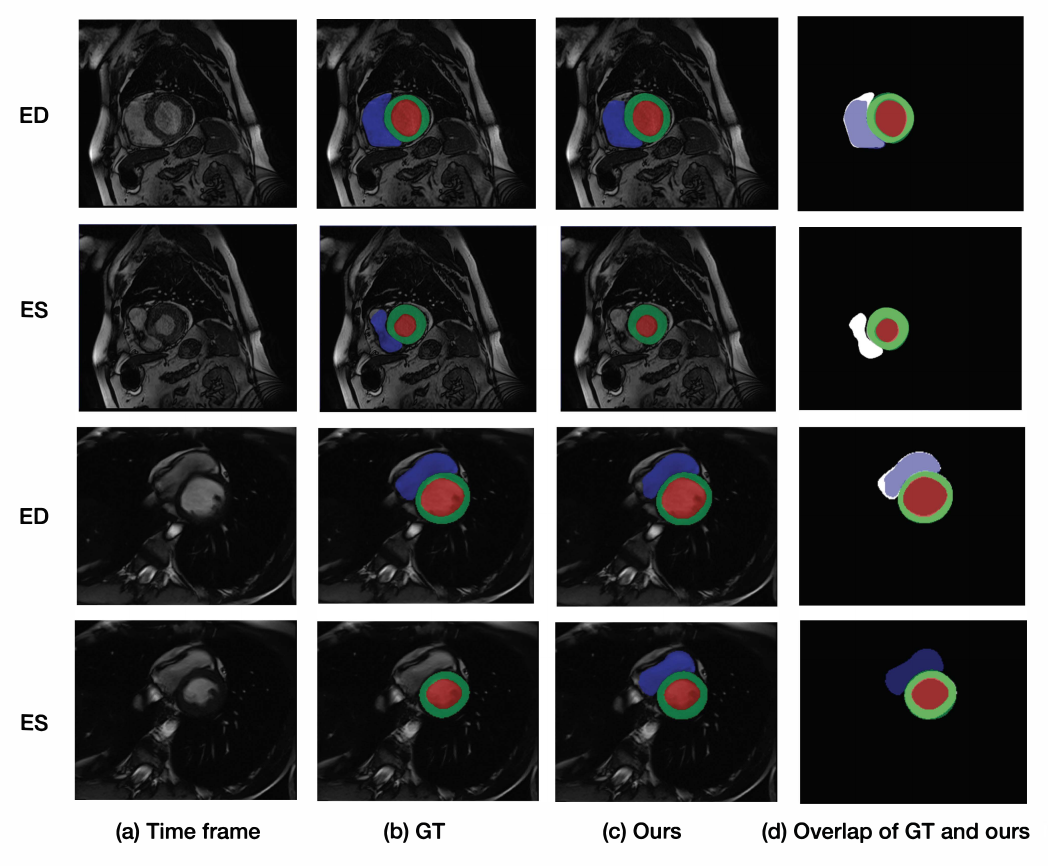}
\caption{Visual examples of inaccurate segmentation. From left to right, it shows the timeframe, ground truth, prediction of the proposed approach, and the difference between ground truth and prediction. LV, MYO, and RV are marked in red, green, and blue, respectively.}
\label{bad}
\end{figure}

\section{Experiments and Results}
\subsection{Dataset}
The challenge cohort is composed of 350 patients with hypertrophic and dilated cardiomyopathies as well as healthy subjects. All subjects were scanned in clinical centres in three different countries (Spain, Germany and Canada) using four different magnetic resonance scanner vendors (Siemens, General Electric, Philips and Canon).

The training set will contain 150 annotated images from two different MRI vendors (75 each) and 25 unannotated images from a third vendor. The CMR images have been segmented by experienced clinicians from the respective institutions, including contours for the LV and RV blood pools, as well as for the left ventricular MYO.

The 200 test cases correspond to 50 new studies from each of the vendors provided in the training set and 50 additional studies from a fourth unseen vendor, that will be tested for model generalizability. 20\% of these datasets will be used for validation and the rest will be reserved for testing and ranking participants.

\subsection{Implementation Details}
Our method is built upon nnUNet~\cite{isensee2019nnu}, a powerful implementation and out-of-the-box tool for medical image segmentation. Then, we use ANTs~\cite{Avants2009Advanced} as the implementation of label propagation algorithm and using the default parameters. The transform method in the registration is three-stage, including rigid, affine, and deformable SyN transform. As for histogram matching, we utilize the scikit-image~\cite{walt2014scikit}. 

\subsection{Results}
We first evaluate our method on the training set of 150 images, where 120 images are used for training and the rest for validation. Dice similarity is employed for evaluation.
\begin{table}[h]
\centering
\setlength{\tabcolsep}{7mm}{
\begin{tabular}{c|c|c|c|c}
\toprule
\multicolumn{1}{c|}{\multirow{2}{*}{Method}} & \multicolumn{3}{c}{Dice Similarity [\%]}                                                     \\ \cline{2-5} 
\multicolumn{1}{c|}{}                        & \multicolumn{1}{c}{LV} & \multicolumn{1}{c}{MYO} & \multicolumn{1}{c|}{RV} & \multicolumn{1}{c}{Average} \\
\midrule
Baseline  &92.88   &86.48   &87.78 &89.05\\
LP  &92.43   &86.73   &89.59 &89.58\\
LP+HM  &92.46   &87.00   &90.94 &90.13\\
\bottomrule
\end{tabular}
}
\caption{Results on training set}
\label{tab:train}
\end{table}

Table~\ref{tab:train} collects the results. Baseline is the fully supervised method (i.e., the 3D UNet) trained solely on end-systole and end-diastole time-frames for which manual segmentations were available. ``LP'' and ``HM'' denote the proposed semi-supervised method that exploits label propagation and histogram matching. It is observed that label propagation excels baseline by 0.53\% and histogram matching further increases 0.55\% in terms of Dice per case.

\begin{table}[h]
\centering
\setlength{\tabcolsep}{1mm}{
\begin{tabular}{c|c|c|c|c|c|c}
\toprule
\multicolumn{1}{c|}{\multirow{2}{*}{Method}} & \multicolumn{3}{c|}{Dice Similarity [\%]}                                                  & \multicolumn{3}{c}{Hausdorff Distance [mm]}                                                    \\ \cline{2-7} 
\multicolumn{1}{c|}{}                        & \multicolumn{1}{c}{LV} & \multicolumn{1}{c}{MYO} & \multicolumn{1}{c|}{RV} & \multicolumn{1}{c}{LV} & \multicolumn{1}{c}{MYO} & \multicolumn{1}{c}{RV} \\
\midrule
Baseline  &91.15$\pm$7.35   &86.00$\pm$4.28   &88.26$\pm$6.59 &8.83$\pm$4.27 &13.08$\pm$12.19 &12.70$\pm$6.72\\
Proposed  &91.75$\pm$7.09   &86.58$\pm$4.25   &89.32$\pm$6.20 &8.10$\pm$4.27 &12.66$\pm$12.30 &11.39$\pm$6.12\\
\bottomrule
\end{tabular}
}
\caption{Results on validation set}
\label{tab:val}
\end{table}

We then validate our method on the validation set, which is held out by the organizers. The docker image with our method is submitted to the organizers to get the results. As shown in Table~\ref{tab:val}, our method consistently outperforms baseline on LV, MYO and RV. Specifically, it improves 0.75\% in terms of Dice per case, and reduces 0.82mm in terms of Hausdorff Distance on average of the three targets. The results on different vendors are collected in Table~\ref{tab:val_vendor}.

\begin{table}[h]
\centering
\setlength{\tabcolsep}{3.3mm}{
\begin{tabular}{c|c|ccc|ccc}
\toprule
\multirow{2}{*}{Method} & \multicolumn{1}{c|}{\multirow{2}{*}{Vendor}} & \multicolumn{3}{c|}{Dice [\%]}                                                  & \multicolumn{3}{c}{HD [mm]}                                                    \\ \cline{3-8} 
                        & \multicolumn{1}{c|}{}                        & \multicolumn{1}{c}{LV} & \multicolumn{1}{c}{MYO} & \multicolumn{1}{c|}{RV} & \multicolumn{1}{c}{LV} & \multicolumn{1}{c}{MYO} & \multicolumn{1}{c}{RV} \\ \midrule
Baseline                & A                                            & 91.70                  & 86.31                   & 87.46                   & 10.05                  & 11.28                   & 14.21                  \\ \cline{2-8} 
                        & B                                            & 94.43                  & 88.79                   & 92.90                   & 7.31                   & 11.16                   & 8.55                   \\ \cline{2-8} 
                        & C                                            & 88.05                  & 86.02                   & 86.26                   & 9.25                   & 12.07                   & 13.16                  \\ \cline{2-8} 
                        & D                                            & 90.39                  & 82.74                   & 86.34                   & 8.73                   & 18.06                   & 14.99                  \\ \midrule
Proposed                & A                                            & 91.80                  & 86.53                   & 87.95                   & 9.26                   & 11.47                   & 13.58                  \\ \cline{2-8} 
                        & B                                            & 94.64                  & 89.07                   & 93.78                   & 6.50                   & 10.25                   & 7.39                   \\ \cline{2-8} 
                        & C                                            & 89.58                  & 87.20                   & 88.12                   & 8.83                   & 12.35                   & 11.32                  \\ \cline{2-8} 
                        & D                                            & 91.00                  & 83.54                   & 87.44                   & 7.80                   & 16.57                   & 13.26                  \\ \bottomrule
\end{tabular}
}
\caption{Results of different vendors on validation set}
\label{tab:val_vendor}
\end{table}

\begin{table}[h]
\centering
\setlength{\tabcolsep}{7mm}{
\begin{tabular}{c|c|c|c}
\toprule
\multicolumn{1}{c|}{\multirow{2}{*}{Method}} & \multicolumn{3}{c}{Dice Similarity [\%]}                                                     \\ \cline{2-4} 
\multicolumn{1}{c|}{}                        & \multicolumn{1}{c}{LV} & \multicolumn{1}{c}{MYO} & \multicolumn{1}{c}{RV} \\
\midrule
Peter M. Full  & 91.0   & 84.9   & 88.4 \\
Yao Zhang (Ours)  & 90.6   & 84.0   & 87.8 \\
Jun Ma  & 90.2	   & 83.5   & 87.4 \\
Mario Parreño  & 91.2   & 83.8   & 85.3 \\
Fanwei Kong  & 90.2   & 82.8   & 85.7 \\
\bottomrule
\end{tabular}
}
\caption{Results on test set}
\label{tab:test}
\end{table}

\begin{table}[h]
\centering
\setlength{\tabcolsep}{0.8mm}{
\begin{tabular}{c|c|c|c|c|c|c}
\toprule
\multicolumn{1}{c|}{\multirow{2}{*}{Vendor}} & \multicolumn{3}{c|}{Dice Similarity [\%]}                                                  & \multicolumn{3}{c}{Hausdorff Distance [mm]}                                                    \\ \cline{2-7} 
\multicolumn{1}{c|}{}                        & \multicolumn{1}{c}{LV} & \multicolumn{1}{c}{MYO} & \multicolumn{1}{c|}{RV} & \multicolumn{1}{c}{LV} & \multicolumn{1}{c}{MYO} & \multicolumn{1}{c}{RV} \\
\midrule
A  &91.87$\pm$5.92   &84.83$\pm$4.37   &88.47$\pm$5.90 &10.72$\pm$16.32 &11.85$\pm$18.76 &12.46$\pm$9.90\\
\hline
B  &91.58$\pm$7.36   &87.24$\pm$4.58   &88.65$\pm$8.10 &7.85$\pm$3.72 &10.13$\pm$3.72 &11.41$\pm$5.82\\
\hline
C  &89.87$\pm$7.32   &83.38$\pm$6.22   &87.65$\pm$6.36 &7.97$\pm$4.80 &9.97$\pm$4.58 &10.71$\pm$4.76\\
\hline
D  &90.29$\pm$5.54   &82.67$\pm$4.30   &87.07$\pm$10.29 &11.16$\pm$19.01 &13.11$\pm$19.97 &16.03$\pm$21.20\\
\hline
Overall  &90.90$\pm$6.61   &84.53$\pm$5.21   &87.96$\pm$7.84 &9.42$\pm$12.92 &11.85$\pm$14.07 &12.65$\pm$12.40\\
\bottomrule
\end{tabular}
}
\caption{Detailed results of our method on 4 vendors of the test set}
\label{tab:test_details}
\end{table}

We also submit our method as docker image to the organizers for online test~\footnote{The test results are presented by organizers at https://www.ub.edu/mnms/}. Please note that neither post-processing nor ensemble strategy is employed in our evaluation. Table~\ref{tab:test} shows the top $5$ teams and our method ranks $2$nd place, demonstrating the effectiveness of our method for cardiac segmentation from multiple vendors and centers. Table~\ref{tab:test_details} presents the detailed results of our method on the images from 4 different vendors. It is observed that the proposed method obtains consistently promising results on both seen and unseen vendors. Fig.~\ref{good} and Fig.~\ref{bad} show some accurate and inaccurate predictions generated by our method.

\section{Conclusion}
In this paper, we design and develop a semi-supervised method for cardiac image segmentation from multiple vendors and medical centers. We exploit label propagation and iterative refinement to leverage unlabelled data in a semi-supervised manner. We further reduce distribution gap between MRI images from different vendors and centers by histogram matching. The results show that our framework is able to achieve superior performance for robust LV, MYO, and RV segmentation. The proposed method ranks $2$nd place among $14$ competitive teams in the M\&M Challenge.

\bibliographystyle{splncs04}
\bibliography{samplepaper.bib}

\begin{thebibliography}{1}
\providecommand{\url}[1]{\texttt{#1}}
\providecommand{\urlprefix}{URL }
\providecommand{\doi}[1]{https://doi.org/#1}

\bibitem{Avants2009Advanced}
Avants, B.B., Tustison, N., Song, G.: Advanced normalization tools (ants). Or
  Insight  \textbf{1–35} (2009)

\bibitem{victor_m_campello_2020_3715890}
Campello, V.M., Palomares, J.F.R., Guala, A., Marakas, M., Friedrich, M.,
  Lekadir, K.: {Multi-Centre, Multi-Vendor \& Multi-Disease Cardiac Image
  Segmentation Challenge} (Mar 2020). \doi{10.5281/zenodo.3715890},
  \url{https://doi.org/10.5281/zenodo.3715890}

\bibitem{chen2019unsupervised}
{Chen}, C., {Ouyang}, C., {Tarroni}, G., {Schlemper}, J., {Qiu}, H., {Bai}, W.,
  {Rueckert}, D.: Unsupervised multi-modal style transfer for cardiac mr
  segmentation. arXiv preprint arXiv:1908.07344 pp. 209--219 (2019)

\bibitem{chen2020deep}
{Chen}, C., {Qin}, C., {Qiu}, H., {Tarroni}, G., {Duan}, J., {Bai}, W.,
  {Rueckert}, D.: Deep learning for cardiac image segmentation: A review.
  Frontiers in Cardiovascular Medicine  \textbf{7}, ~25 (2020)

\bibitem{chen2020leveraging}
{Chen}, L.C., {Lopes}, R.G., {Cheng}, B., {Collins}, M.D., {Cubuk}, E.D.,
  {Zoph}, B., {Adam}, H., {Shlens}, J.: Leveraging semi-supervised learning in
  video sequences for urban scene segmentation. arXiv preprint arXiv:2005.10266
   (2020)

\bibitem{isensee2019nnu}
{Isensee}, F., {Petersen}, J., {Kohl}, S.A.A., {Jäger}, P.F., {Maier-Hein},
  K.H.: nnu-net: Breaking the spell on successful medical image segmentation.
  arXiv preprint arXiv:1904.08128  (2019)

\bibitem{ronneberger2015u}
Ronneberger, O., Fischer, P., Brox, T.: U-net: Convolutional networks for
  biomedical image segmentation. In: International Conference on Medical Image
  Computing and Computer-Assisted Intervention. pp. 234--241. Springer (2015)

\bibitem{tao2019deep}
Tao, Q., Yan, W., Wang, Y., Paiman, E.H., Shamonin, D.P., Garg, P., Plein, S.,
  Huang, L., Xia, L., Sramko, M., et~al.: Deep learning--based method for fully
  automatic quantification of left ventricle function from cine mr images: a
  multivendor, multicenter study. Radiology  \textbf{290}(1),  81--88 (2019)

\bibitem{walt2014scikit}
van~der {Walt}, S., {Schönberger}, J.L., {Nunez-Iglesias}, J., {Boulogne}, F.,
  {Warner}, J.D., {Yager}, N., {Gouillart}, E., {Yu}, T.: scikit-image: Image
  processing in python. PeerJ  \textbf{2}(1) (2014)

\end{thebibliography}
\end{document}